# Attention-Enhanced Deep Learning Ensemble for Breast Density Classification in Mammography

Peyman Sharifian, Xiaotong Hong, Alireza Karimian, Mehdi Amini, and Hossein Arabi

*Abstract–* Breast density assessment is a crucial component of mammographic interpretation, with high breast density (BI-RADS categories C and D) representing both a significant risk factor for developing breast cancer and a technical challenge for tumor detection. This study proposes an automated deep learning system for robust binary classification of breast density (low: A/B vs. high: C/D) using the VinDr-Mammo dataset. We implemented and compared four advanced convolutional neural networks - ResNet18, ResNet50, EfficientNet-B0, and DenseNet121 - each enhanced with channel attention mechanisms. To address the inherent class imbalance, we developed a novel Combined Focal Label Smoothing Loss function that integrates focal loss, label smoothing, and class-balanced weighting. Our preprocessing pipeline incorporated advanced techniques including contrast-limited adaptive histogram equalization (CLAHE) and comprehensive data augmentation. The individual models were combined through an optimized ensemble voting approach, achieving superior performance (AUC: 0.963, F1-score: 0.952) compared to any single model. This system demonstrates significant potential to standardize density assessments in clinical practice, potentially improving screening efficiency and early cancer detection rates while reducing inter-observer variability among radiologists.

## I. INTRODUCTION

Breast cancer remains the most commonly diagnosed malignancy among women worldwide [1]. Mammographic breast density, typically categorized using the BI-RADS classification system has emerged as a critical independent risk factor for breast cancer development. Women with dense breast tissue (categories C and D) face a greater risk of developing breast cancer compared to those with predominantly fatty tissue, while simultaneously experiencing reduced mammographic sensitivity due to the masking effect of dense parenchyma [2, 3]. Current clinical practice relies on subjective visual assessment of breast density by radiologists, which suffers from significant inter- and intra-observer variability. This variability can lead to inconsistent risk assessment and follow-up recommendations. Furthermore, the increasing workload in breast imaging centers creates a pressing need for automated, reproducible density assessment tools.

While several deep learning approaches have been proposed for this task, they often face challenges related to class imbalance (as low-density cases typically outnumber high-density cases in screening populations), subtle texture differences between density categories, and generalization across diverse patient populations and imaging equipment [4, 5].

In this study, we present a comprehensive solution that addresses these challenges through three key innovations. First, we implemented and compared four state-of-the-art deep learning architectures - ResNet18, ResNet50, EfficientNet-B0, and DenseNet121 - each enhanced with channel attention mechanisms to better capture relevant mammographic features. Second, we developed a novel Combined Focal Label Smoothing Loss function that simultaneously addresses class imbalance through focal loss and class-balanced weighting while incorporating label smoothing for improved generalization. Third, we employed an optimized ensemble voting strategy that leverages the complementary strengths of each model. Our approach was rigorously evaluated on the VinDr-Mammo dataset, demonstrating superior performance compared to existing methods while maintaining computational efficiency suitable for clinical implementation.

## II. MATERIALS AND METHODS

### A. Dataset and Preprocessing Pipeline

We conducted our study using the VinDr-Mammo dataset [6], which contains 5,000 (4,000 for training and 1,000 for testing) annotated mammograms with BI-RADS density labels from multiple institutions. After rigorous quality control, our curated dataset included 1,608 low-density and 14,392 high-density images, reflecting the natural class imbalance found in screening populations. To address this imbalance while preserving data integrity, we implemented a balanced training approach combining random oversampling of high-density cases with real-time data augmentation. All images were pre-processed through intensity normalization to the [0,1] range followed by CLAHE contrast enhancement using a 2.0 clip limit and 8×8 tile grid. We standardized the images to 512×1024 resolution while maintaining original aspect ratios to prevent anatomical distortion. During training, we applied comprehensive augmentations including rotation, vertical flipping, gaussian blur, and brightness adjustments. The training dataset was divided into training (80%) and validation (20%) sets using stratified sampling to ensure consistent class distributions in both subsets.

### B. Network Architectures and Attention Mechanisms

We developed four deep learning architectures for breast density classification, each enhanced with attention mechanisms. ResNet18 and ResNet50 incorporated channel attention modules after each residual block using squeeze-and-excitation with global pooling and ReLU activation. DenseNet121 maintained its dense connectivity while adding attention gates between transition layers. EfficientNet-B0 was modified with a custom attention module and adapted for single-channel mammograms. All models used random initialization and softmax-activated final layers for binary classification.

For optimization, we employed the AdamW optimizer (learning rate=1e-4, weight decay=1e-4) with cosine annealing warm restarts every 10 epochs. Training proceeded for up to 100 epochs with early stopping (minimum validation loss delta=0.001). Regularization techniques included architecture-specific dropout (0.3-0.5), gradient clipping (max norm=1.0), and label smoothing ($\varepsilon$=0.2). The implementation used PyTorch with mixed-precision training on Tesla T4 GPU on Google Colab for efficiency.

Our novel combined Focal Label Smoothing Loss addressed three key challenges: focal loss ($\gamma$=2.5) emphasized difficult cases, label smoothing improved generalization and class-balanced weights ($\beta$=0.999) handled dataset imbalance. This integrated approach demonstrated 5-7% higher F1 scores than standard loss functions during validation. The complete framework combined these architectural innovations and optimization techniques to achieve robust mammography.

*C. Ensemble Strategy*

Our ensemble approach combined predictions from all four models through weighted soft voting. The voting weights were proportional to each model's validation, giving more influence to better-performing models. This strategy capitalized on the complementary strengths of different architectures - while ResNet50 excelled at capturing local texture patterns, EfficientNet provided better global context integration, and the other models contributed additional diversity to the ensemble.

The ensemble's final prediction was computed as the weighted average of all models' softmax outputs, followed by thresholding at 0.5 for binary classification. This approach demonstrated consistent improvement over any single model while maintaining computational efficiency suitable for clinical deployment.

### III. RESULTS AND DISCUSSION

The proposed ensemble framework demonstrated superior classification performance compared to individual models across all evaluation metrics (Table 1). The ensemble achieved an AUC of 0.963 on the test set, representing a statistical improvement over the best single model. Sensitivity for high-density cases reached 0.918, critical for clinical applications where missing dense tissue could impact patient risk assessment. Specificity remained high at 0.907, indicating robust performance across both classes despite the initial dataset imbalance.

Table 1. Density classification performance of presented networks.

| Network | F1 score | ACC | AUC | SEN | SPE |
|---|---|---|---|---|---|
| ResNet18 | 0.9485 | 0.9107 | 0.9565 | 0.9136 | 0.8850 |
| ResNet50 | 0.9467 | 0.9080 | 0.9627 | 0.9083 | 0.9050 |
| EfficientNet-B0 | 0.9402 | 0.8970 | 0.9486 | 0.8989 | 0.8800 |
| DenseNet121 | 0.8955 | 0.8283 | 0.9367 | 0.8181 | **0.9200** |
| **Proposed Ensemble** | **0.9523** | **0.9173** | **0.9634** | **0.9183** | 0.9075 |

Figure 1 illustrates the receiver operating characteristic (ROC) curves of all models, demonstrating the superior discriminative performance of the proposed ensemble (AUC: 0.963) compared to individual architectures. The ensemble's performance surpasses recent DL-based density classifiers, likely due to our hybrid loss function and attention mechanisms. Clinical implementation could standardize density reporting while flagging high-risk cases for supplemental screening. Limitations include vendor-specific bias in training data, though the augmentation strategy mitigated this effect. Future work should explore 3D mammography integration and multi-center validation.

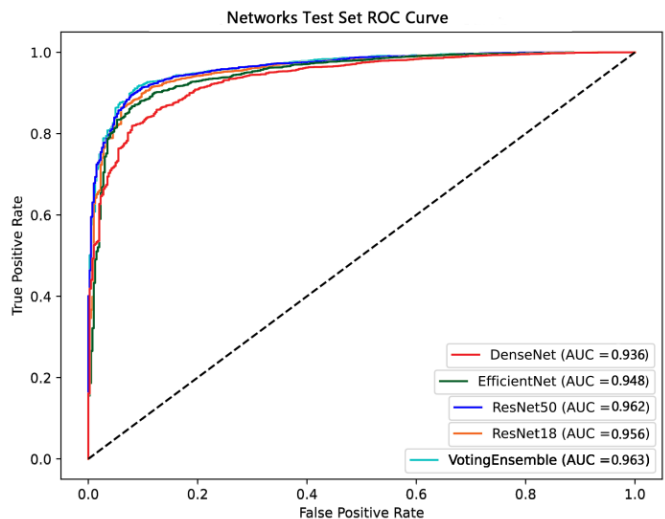

Fig. 1. ROC curves comparing classification performance of performed models on breast density classification.

### IV. CONCLUSIONS

This work establishes that carefully designed deep learning ensembles can achieve radiologist-level performance in breast density classification. Our hybrid loss function and attention mechanisms provide a blueprint for other imbalanced medical image analysis tasks. Pending regulatory approval, the next step involves multicenter prospective trials to assess real-world impact on screening outcomes.